\documentclass[%
 reprint,
superscriptaddress,
prb,
]{revtex4-2}
\usepackage{graphicx}
\usepackage{dcolumn}
\usepackage{bm}
\usepackage{float}
\usepackage{array}
\usepackage{booktabs}
\usepackage{amssymb}
\usepackage{bbding}
\usepackage{amsmath}
\usepackage[dvipsnames]{xcolor}

\usepackage[colorlinks=true,linkcolor=blue,anchorcolor=red,%
citecolor=blue,urlcolor=blue]{hyperref}

\usepackage[mathlines]{lineno}

\begin{document}

\title{Complete Magnetic Hierarchy in Bichromatically Driven
Unconventional Magnets
}

\author{Pei-Hao Fu}
\email{peihao.fu@gmail.com}
\address{Department of Physics and Astronomy, University of
Florence, I-50019 Sesto Fiorentino, Italy}

\author{Sayed Ali Akbar Ghorashi}
\email{ghorashi@sas.upenn.edu}
\address{Department of Chemistry, University of Pennsylvania,
Philadelphia, USA}

\date{\today}

\begin{abstract}
We establish a general framework for engineering unconventional magnetism using bichromatic driving.
We identify a sharp hierarchy: for a bichromatic drive with a generic even commensurate frequency ratio, the leading continuum $q$-wave magnet generates components only up to order $q-2$, whereas lattice higher-gradient terms supply the missing $q-1$ and $q$ parities and complete the hierarchy through order $q$. 
We explicitly illustrate this new revised hierarchy on a $d$-wave altermagnet irradiated by a circularly-linearly polarized light, and show that the asymmetry of the driving field is transferred to the light-dressed electronic structure, producing momentum-asymmetric dispersions beyond the static and single-frequency limits. 
We further show that this band asymmetry is absent in the leading continuum expansion and originates from higher-gradient lattice terms that are converted by the bichromatic drive into odd-in-momentum contributions. 
Extending the analysis to higher-order unconventional magnets, we further obtain complete magnetic hierarchies for all $q$-parity unconventional magnets. 
Our results establish bichromatic Floquet driving as a general route for generating mixed-parity magnetic hierarchies and dynamically engineering unconventional magnetism.
\end{abstract}
\maketitle

Unconventional magnets (UMs) provide a promising platform for quantum phenomena beyond conventional magnetism due to their distinct magnetic parities, rooted in the interplay between spin and crystalline symmetries
\cite{Smejkal2022a,Smejkal2022,Yuri2025Superconducting,liu2025review}.
An increasing number of novel phenomena have been discovered in UMs, including multipolar quantum-geometric distributions \cite{ghorashiALMnonlinear,5bns-qkgl},
topological phases \cite{ghorashiALMTSC,TSCALM2,
PhysRevB.111.L121401,bfwd-hcm5,jiang2026altermagnetism,mcardle2026topologicalsuperconductivityaltermagneticheterostructures,Zhu2023Topological,Chatterjee2025Interplay,Li2023Majorana,Li2024Realizing,Tanaka2024Theory,Chen2026Topological},
spintronic functionalities \cite{Bai2024Altermagnetism,Fu2025All,Yuan2026,Liu2026Intrinsic,fu2026perfectspinnonreciprocitygated,
ifmmode2022Giant,Bai2022Observation,Karube2022Observation,
Han2024Harnessing,Samanta2024Spin,Sun2023Spin,fu2026universaltransversediodeeffects,Chen2026NeelVector,Gong2026Defect}, superconductivity \cite{Maeda2025Classification,Fukaya2025Josephson,Chakraborty2024Constraints,Sukhachov2024Coexistence,Sun2023Andreev,Papaj2023Andreev, Nagae2025Spin,Zhao2025Orientation,Niu2024Orientation,Maeda2024Tunneling,Das2024Crossed,Nagae2025Spin,Niu2024Electrically,Fukaya2025Josephson,Zhang2024Finite,Lu2024Josephson,Sun2025Tunable,Ouassou2023dc,Beenakker2023Phase,Banerjee2024Altermagnetic,Cheng2024FieldFree,Chakraborty2024Perfect,gkrn-pbjb,Tolbatov2026Altermagnetism,Kia2026TwistTunable,Niebuhr2026SpinPolarized}
and a broad range of emergent phenomena
\cite{Latorre2026Altermagnetic,Tagani2026Band,Tagani2026RankSelective,Flynn2026Hidden,Gobel2026Orbital,MPA_ALM,Smejkal2022c,Maznichenko2024Fragile,
Lin2025Coulomb,Chen2024Electrical,Reja2024,
GonzalezBetancourt2023,Dufouleur2023,Reichlova2024,Feng2022}, 
together with an increasing number of material realizations
\cite{Liu2024c,Feng2022,Dufouleur2023,Hariki2024,
GonzalezBetancourt2023,Reichlova2024,Rial2024,Ding2024,
Yu2025Neel,Li2025Altermagnetism,Faure2025Altermagnetism,
Elmers2020}.
These rich phenomena can be largely attributed to the
momentum-dependent spin splitting constrained by combined spin and spatial symmetries, making its controlled manipulation a central question for UMs.

\begin{figure}[t!]
    \centering
    \includegraphics[width=0.9\linewidth]{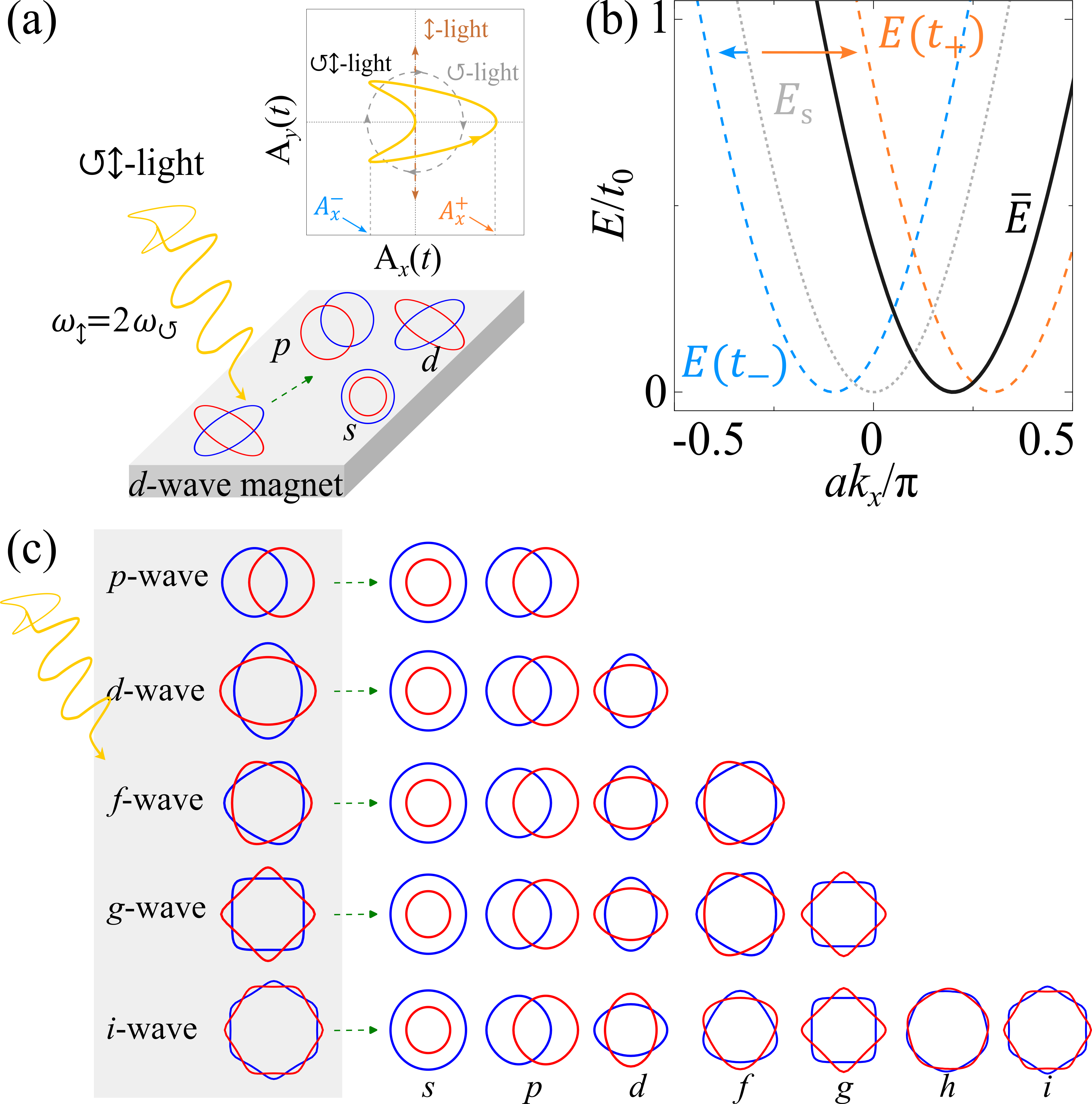}
    \caption{
    (a) Schematic of a $d$-wave magnet illuminated by a two-frequency driving field composed of circularly ($\circlearrowleft$) and linearly ($\updownarrow$) polarized light, with the corresponding vector-potential trajectory in
    the $A_x(t)$-$A_y(t)$ plane of individual and combined lights.
    (b) Schematic band dispersions of static state, $E_s$, instantaneous band dispersions at $t_{\pm}$, $E(t_{\pm})$, and $\bar{E}$ after time averaging.
    (c) Summary of results: Light-induced hierarchy of spin-splitting components generated by the bichromatic drive from parent $p$-, $d$-, $f$-, $g$-, and $i$-wave magnets.
    }
    \label{figure1}
\end{figure}

Floquet engineering provides a powerful framework for dynamically manipulating electronic structures
\cite{Oka2019,rudner2020band,weitenberg2021tailoring}. 
Despite growing interest in monochromatic
\cite{Ghorashi2025Dynamical,
Farajollahpour2025Light,tian2026opticallydrivenorbitalhall,
cheraghchi2026floquettopologicalphasesanomalous,
maryam2026lightinducedtopologicalphasetransitions,
Fu2025Light,Fu2025Floquet,yokoyama2025FloquetSC,Feng2026Coupled,Li2026Dynamical,Han2026Optical,Zhang2026QuantumMetric}
and bichromatic Floquet engineering
\cite{Ikeda2022a,Ikeda2023,Nag2019,Arakawa2024,Trevisan2022,
Ganguli2025,
17rp-cx9f,ganguli2025tunabletopologyhallresponse,
yu2026tunableoddparityspinsplittings,Zhu2026Floquet,17rp-cx9f},
a fundamental understanding of how time-dependent driving fields break the symmetries that protect unconventional magnetism remains lacking.
This question is particularly relevant for bichromatic driving, which can break spatial symmetries \cite{Ikeda2022a} in addition to time-reversal symmetry, precisely the symmetries that constrain UMs.
Recent work has proposed bichromatically induced odd-parity spin splitting in altermagnets
\cite{yu2026tunableoddparityspinsplittings}.
Here we address a different question: \textit{can light turn a single $q$-wave magnet into a complete hierarchy of magnetic orders, from $s$-wave up to $q$-wave, and what does it take?}
Here, a $q$-wave order denotes a spin splitting whose leading momentum dependence is of order $k^{q}$, with $q=1,2,3,4,6$ for $p$-, $d$-, $f$-, $g$-, and $i$-wave, respectively.

\begin{figure}
    \centering
    \includegraphics[width=1\linewidth]{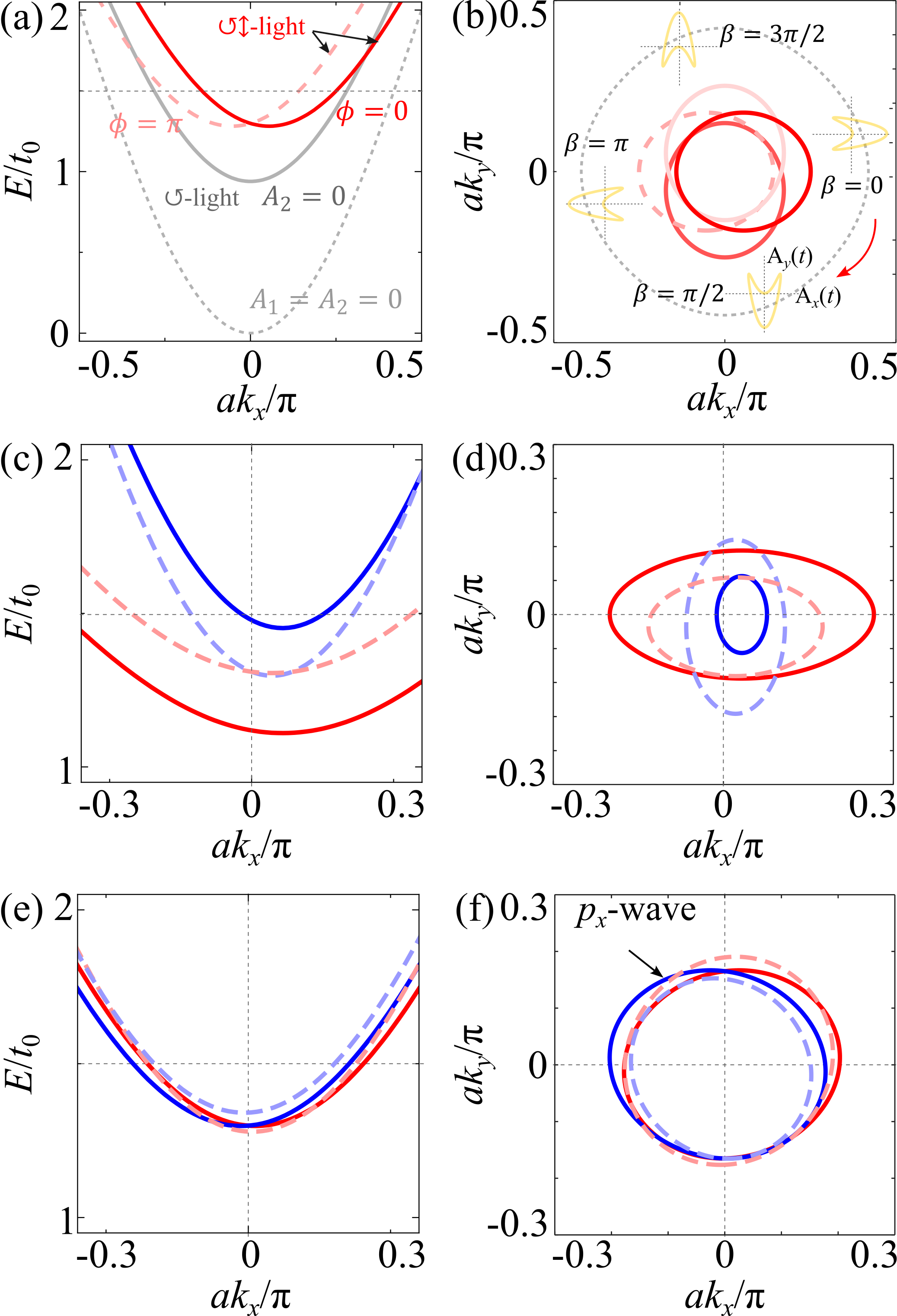}
    \caption{
    (a) Dispersion of Eq.\,(\ref{HeffDecomp}) for $t_J=0$, $k_y=0$, and $\beta=0$.
    (b) Equal-energy contours at $E/t_0=1.5$ for different $\beta$ with $\phi=0$.
    (c) Two-frequency-light-dressed dispersion of the $d_{x^{2}-y^{2}}$ magnet ($\theta_{J}=0$) for $\beta=\pi/4$ (dashed lines) and $\beta=0$ (solid lines).
    The spin-up (spin-down) states are denoted in red (blue).
    (d) Corresponding equal-energy contours at $E/t_{0}=1.5$.
    (e,f) Same as (c,d), but for the $d_{xy}$ magnet
    ($\theta_{J}=\pi/4$) and $\phi=\pi/2$.
    Parameters: $t_0=1$, $\omega=1$, $\eta=1$, $\mu=0$, $a e A_1/\hbar=a e A_2/\hbar=1$, and $t_{J}=0$ for (a,b) and $t_{J}=0.5$ for (c-f).
    }
    \label{figure2}
\end{figure}

We show that, in leading continuum theory, a parent $q$-wave magnet acquires light-induced components only up to order $q-2$, whereas the full lattice Hamiltonian generates all induced components up to order $q$. This completeness requires two microscopic ingredients: the higher-gradient terms of the lattice spin splitting, and an even commensurate frequency ratio of the bichromatic drive. 
Together, they activate the otherwise missing components of orders $q-1$ and $q$.
Our results are summarized in Fig.\,\ref{figure1}(c).
We first derive this rule for a generic lattice spin splitting, showing how the asymmetry of the driving field is transferred to the light-dressed electronic structure (Fig.\,\ref{figure1}(a,b)).
We then illustrate it with a $d$-wave altermagnet driven by a circularly-linearly polarized light, which acquires a complete $s$-$p$-$d$ hierarchy (Fig.\,\ref{figure2}), and with an odd-parity $f$-wave magnet; the $g$- and $i$-wave cases are given in the \textbf{End Matter}.

\textit{Floquet hierarchy of unconventional magnetic orders.}---
We consider a two-dimensional collinear lattice
magnet, $H_{\sigma}(\bm{k})=\xi(\bm{k})+\sigma M^{q}_{\mathrm{lat}}(\bm{k})$,
where 
$\xi(\bm{k})$ is the
kinetic energy, $M^{q}_{\mathrm{lat}}(\bm{k})
=\sum_{\bm{R}}m^{q}_{\bm{R}}e^{i\bm{k}\cdot\bm{R}}$ is $q$-parity spin splitting given by a sum of harmonics, with lattice bonds $\bm{R}$, and $\sigma=\pm1$ labels the spin projection, driven by a two-frequency field composed of circularly ($\circlearrowleft$)
and linearly ($\updownarrow$) polarized components,
\begin{equation}
\bm{A}(t)
=
\bm{A}_{1}^{\circlearrowleft}(t)
+
\bm{A}_{2}^{\updownarrow}(t),
\label{Atotal}
\end{equation}
with
\begin{subequations}
\label{Afields}
\begin{align}
\bm{A}_{1}^{\circlearrowleft}(t)
={}&
A_{1}[
\cos(\omega_1 t-\phi_{1}),
\eta\sin(\omega_1 t-\phi_{1})
],
\\
\bm{A}_{2}^{\updownarrow}(t)
={}&
A_{2}
\cos(\omega_2 t-\phi_{2})
\left(
\cos\beta,\sin\beta
\right),
\end{align}
\end{subequations}
where $A_{1,2}$ and $\phi_{1,2}$ denote the amplitudes and
initial phases of the two fields, respectively; $\eta=\pm1$
determines the handedness of the $\circlearrowleft$ component,
while $\beta$ specifies the polarization direction of the
$\updownarrow$ component.
For the $1{:}2$ drive we define the phase combination
$\phi\equiv2\phi_{1}-\phi_{2}$.
Hereafter, we set $\omega_{1}\equiv\omega$ and focus on
$\mathcal{N}=\omega_{2}/\omega_{1}=2$, so that $\omega_{2}=2\omega$ and the
total drive has period $T=2\pi/\omega$.
As illustrated in Fig.\,\ref{figure1}(a), the superposition of
the $\omega$-$\circlearrowleft$-light and the
$2\omega$-$\updownarrow$ one traces an asymmetric trajectory in
the $A_{x}$-$A_{y}$ plane, manifested by, e.g., the unequal
positive and negative excursions of the $x$ component,
$A_{x}^{+}(t_+)\neq-A_{x}^{-}(t_-)$, where
$A_{x}^{+(-)}=\max_{t}[\min_{t}]A_{x}(t)$ corresponds to
$t_{+(-)}$.
After the Peierls substitution,
$H_{\sigma}(\bm{k})$ becomes time dependent as
$H_{\sigma}[\bm{k}+e\bm{A}(t)/\hbar]$, and the instantaneous
dispersion $E_{\sigma}(\bm{k},t)$ acquires different displacement
magnitudes in opposite momentum directions at $t_{+}$ and $t_{-}$
due to $A_{x}^{+}\neq-A_{x}^{-}$, see Fig.\,\ref{figure1}(b).
Thus, after evolution over one complete driving period, the time-averaged dispersion becomes asymmetric, $\bar{E}_{\sigma}(\bm{k})\neq\bar{E}_{\sigma}(-\bm{k})$
[Fig.\,\ref{figure1}(b)].
This intuitive picture provides a phenomenological understanding of how a two-frequency driving field can generate band asymmetry even when the static system is initially inversion symmetric.
More precisely, odd $\mathcal{N}$ enforces the half-period relation $\bm{A}(t+T/2)=-\bm{A}(t)$ and therefore forbids the odd-in-$\bm{k}$ component, while even $\mathcal{N}$ removes this symmetry protection and allows, but does not by itself guarantee, a finite band asymmetry; see the example given in the \textbf{End Matter}.

The above phenomenological picture can be confirmed rigorously using Floquet theory \cite{Oka2019,rudner2020band,Ikeda2022a}.
Since $H_{\sigma}(\bm{k})$ is spin diagonal,
$H_{\sigma}[\bm{k}+e\bm{A}(t)/\hbar]$ commutes with itself at
arbitrary times, and consequently the stroboscopic Floquet
Hamiltonian is exactly given by the period average,
\begin{equation}
H_{\sigma}^{\mathrm{eff}}(\bm{k})
=
\left\langle
H_{\sigma}[\bm{k}+e\bm{A}(t)/\hbar]
\right\rangle,
\label{Heff-average}
\end{equation}
where $\langle\cdots\rangle=T^{-1}\int_{0}^{T}dt\,(\cdots)$.
Hereafter, the factors $ea/\hbar$ are absorbed into the
dimensionless amplitudes $A_{1,2}$.

To further elucidate the lattice origin of this asymmetry, we compare this with the leading continuum theory. 
Along a fixed momentum direction, denoted by $k_i$ ($i\in\{x,y\}$), an inversion-symmetric band $\xi(k_i)=\alpha k_i^{2}+\gamma k_i^{4}+\cdots$ gives $\langle\xi(k_i+A_i)\rangle=\alpha[k_i^{2}+\langle A_i^{2}\rangle] +\gamma[k_i^{4}+6k_i^{2}\langle A_i^{2}\rangle+4k_i\langle A_i^{3}\rangle +\langle A_i^{4}\rangle]+\cdots$. Since $\langle A_i\rangle=0$, the parabolic continuum band remains even in $k_i$, whereas the first non-parabolic correction generates
the odd term $4\gamma k_i\langle A_i^{3}\rangle$.
For Eq.\,(\ref{Afields}) with $\mathcal{N}=2$, one obtains
$\langle A_x^3\rangle=3A_1^2A_{2x}\cos\phi/4$ and
$\langle A_y^3\rangle=-3A_1^2A_{2y}\cos\phi/4$, showing that the bichromatic field converts higher-gradient lattice structure into a lower-order band asymmetry.

Let us now understand the origin of this hierarchy.
After the Peierls substitution and time averaging, each harmonic of the spin splitting $M^{q}_{\mathrm{lat}}(\bm{k})$ simply picks up a light-dressing factor, \begin{equation}
M^{q}_{\mathrm{eff}}(\bm{k})
=
\sum_{\bm{R}}m^{q}_{\bm{R}}\,\mathcal{F}_{\bm{R}}\,
e^{i\bm{k}\cdot\bm{R}},
\;
\mathcal{F}_{\bm{R}}
=
\left\langle e^{i\bm{A}(t)\cdot\bm{R}}\right\rangle.
\label{Mlattice-dressed}
\end{equation}
The magnitude of $\mathcal{F}_{\bm{R}}$ renormalizes the amplitude of each harmonic, similar to the dynamical localization of the kinetic term \cite{rudner2020band}, while its phase mixes the cosine and sine parts of
the harmonic. 
The latter is the key: it generates an even harmonic from an odd one and vice versa, 
and is therefore responsible for the parity conversion. 
Therefore, expanding Eq.\,(\ref{Mlattice-dressed}) around the $\Gamma$ point gives the $s$-, $p$-, $d$-, $f$-, $g$-, $\ldots$ components to complete the magnetic hierarchy.

This picture shows that two ingredients are needed for a complete hierarchy. 
(i) An even frequency ratio. 
The phase of $\mathcal{F}_{\bm{R}}$ comes from the odd powers of $\bm{A}(t)$, which survive time averaging only when their frequencies cancel.
For odd $\mathcal{N}$, this never happens: $\mathcal{F}_{\bm{R}}$ is real, and light only rescales each harmonic, so the hierarchy keeps the
parity of the parent magnet. 
In contrast, for even $\mathcal{N}$, the lowest odd power that survives the time average is $\langle A^{\mathcal{N}+1}\rangle$, which combines $\mathcal{N}$ photons of the $\omega_1$ beam with one photon of the $\omega_2=\mathcal{N}\omega_1$ beam and therefore scales as $A_1^{\mathcal{N}}A_2$.
(ii) Higher-gradient terms of the lattice spin splitting. 
Suppose we instead start from the leading continuum form of a $q$-wave spin splitting, $M(\bm k)\propto k^{q}$. 
Since the time average of $\bm{A}(t)$ itself vanishes, light can only lower the order of this term by two or more, and it leaves the $k^{q}$ term untouched. 
As a result, the continuum model never induces the $k^{q}$ and $k^{q-1}$ components. 
In contrast, the lattice spin splitting also contains higher-order terms $k^{q+2},k^{q+4},\ldots$, which are dropped in the continuum limit, and light brings them down to orders $q$ and $q-1$, completing the induced hierarchy up to $q$.
For the $\mathcal{N}=2$ drive used here, the continuum induces $s$ and $(s,p)$, for the $d$-, $f$-wave parents, whereas the lattice gives $(s,p,d)$ and $(s,p,d,f)$, respectively [Fig.\,\ref{figure1}(c)].
It is important to note that for larger even $\mathcal{N}$ the continuum misses even more opposite-parity components, all of which are supplied by the lattice, and that special phases or polarizations can make individual components vanish.

In the rest of the Letter, we explicitly demonstrate this for two examples: an even-parity $d$-wave magnet and an odd-parity $f$-wave magnet.

\textit{$d$-wave altermagnet: An even-parity case}---
We start with a $d$-wave altermagnet,
$H_{\sigma}(\bm{k})=\xi(\bm{k})+\sigma M_{\mathrm{lat}}^{(d)}(\bm{k})$
\cite{Smejkal2022a,Smejkal2022,Yuri2025Superconducting}, where
$\xi(\bm{k})=4t_{0}-\mu-2t_{0}(\cos k_{x}+\cos k_{y})$ is the
kinetic energy with nearest-neighbor hopping $t_{0}$ and chemical
potential $\mu$, and $\sigma=\pm1$ labels the spin parallel
($\uparrow$) or antiparallel ($\downarrow$) to the N\'eel vector.
The altermagnetic order is given by
\begin{equation}
M_{\mathrm{lat}}^{(d)}(\bm{k})
=
-2t_{J}^{C}\left(\cos k_{x}-\cos k_{y}\right)
-\frac{t_{J}^{S}}{2}\sin k_{x}\sin k_{y},
\label{Mstatic}
\end{equation}
where $t_{J}^{C}=t_{J}\cos2\theta_{J}$ and
$t_{J}^{S}=t_{J}\sin2\theta_{J}$, $t_{J}$ sets the strength of the UM, and $\theta_{J}=0$ ($\pi/4$) corresponds to the $d_{x^{2}-y^{2}}$ ($d_{xy}$) case.
This order respects the spin-group symmetry $[C_{2}\|C_{4}^{z}]$, enforcing $d$-wave spin splitting \cite{Smejkal2022,Smejkal2022a}.

Following the Floquet procedure described above, the light-induced effective Hamiltonian is obtained as
$H_{\sigma}^{\mathrm{eff}}=\xi^{\prime}+\xi^{\prime\prime}
+\sigma(M^{\prime}+M^{\prime\prime})$, with
\begin{subequations}
\label{HeffDecomp}
\begin{align}
\xi^{\prime}(\bm{k})
={}&
4t_{0}-\mu-2\left(t_{0}^{x}\cos k_{x}+t_{0}^{y}\cos k_{y}\right),
\label{xiprime}
\\
M^{\prime}(\bm{k})
={}&
-2\left(t_{J}^{Cx}\cos k_{x}-t_{J}^{Cy}\cos k_{y}\right)\cr
&-t_{J}^{S+}\sin k_{x}\sin k_{y}
-t_{J}^{S-}\cos k_{x}\cos k_{y},
\label{Mprime}
\\
\xi^{\prime\prime}(\bm{k})
={}&
-2\left(t_{1}^{x}\sin k_{x}-t_{1}^{y}\sin k_{y}\right),
\label{xidoubleprime}
\\
M^{\prime\prime}(\bm{k})
={}&
-2\left(t_{J'}^{Cx}\sin k_{x}+t_{J'}^{Cy}\sin k_{y}\right)\cr
&-t_{J'}^{S+}\sin k_{x}\cos k_{y}
-t_{J'}^{S-}\cos k_{x}\sin k_{y}.
\label{Mdoubleprime}
\end{align}
\end{subequations}
The light-dressed parameters
$t_{0}^{i}\simeq t_{0}J_{0}(A_{1})J_{0}(A_{2i})$,
$t_{J}^{Ci}\simeq t_{J}^{C}J_{0}(A_{1})J_{0}(A_{2i})$, and
$t_{J}^{S\pm}\simeq t_{J}^{S}J_{0}(\sqrt{2}A_{1})
[J_{0}(A_{2-})\pm J_{0}(A_{2+})]/4$ are products of the dressings
by each light separately, whereas the light-induced parameters
$t_{1}^{i}\simeq2t_{0}\cos\phi\,J_{12}^{i}$,
$t_{J'}^{Ci}\simeq2t_{J}^{C}\cos\phi\,J_{12}^{i}$, and
$t_{J'}^{S\pm}\simeq\eta\,t_{J}^{S}\sin\phi\,J_{12}^{\pm}/2$, with
$J_{12}^{i}=J_{1}(A_{2i})J_{2}(A_{1})$ and
$J_{12}^{\pm}=J_{2}(\sqrt{2}A_{1})
[J_{1}(A_{2+})\pm J_{1}(A_{2-})]$, require both lights.
Here $i=x,y$, $J_{n}$ is the Bessel function of the first kind,
$A_{2x}=A_{2}\cos\beta$, $A_{2y}=A_{2}\sin\beta$, and
$A_{2\pm}=A_{2x}\pm A_{2y}$; we retain the leading terms of the
Bessel expansion, and the exact coefficients follow from the bond
factors in Eq.\,(\ref{Mlattice-dressed}).
The two parts directly reflect the two ingredients identified
above.
(i) The even part, $\xi^{\prime}+\sigma M^{\prime}$, only renormalizes the static harmonics and is present for either light alone.
(ii) The odd part, $\xi^{\prime\prime}+\sigma M^{\prime\prime}$, is the genuine frequency-mixing contribution.
It scales as $A_{1}^{2}A_{2}$ and therefore requires both lights
at the even ratio $\mathcal{N}=2$, as well as the $k^{4}$ content
of the lattice harmonics.
Consequently,
$H_{\sigma}^{\mathrm{eff}}(\bm{k})\neq H_{\sigma}^{\mathrm{eff}}(-\bm{k})$,
consistent with Fig.\,\ref{figure1}(b).

We first set $t_{J}=0$ to isolate the band asymmetry.
As shown in Fig.\,\ref{figure2}(a), either light alone only dresses the band through the dynamical localization of $t_{0}^{x,y}$ \cite{rudner2020band}, and the band remains even in $\bm{k}$.
When both lights are present, $\xi^{\prime\prime}$ displaces the band center to $(\delta_{x},-\delta_{y})$, with $\tan\delta_{i}=t_{1}^{i}/t_{0}^{i}\simeq2\cos\phi\,
J_{1}(A_{2i})J_{2}(A_{1})/[J_{0}(A_{1})J_{0}(A_{2i})]$, so that both the magnitude and the direction of the displacement are tuned by $\phi$ and $\beta$ [Fig.\,\ref{figure2}(b)].
Using $\gamma=-t_{0}/12$ from the expansion of $-2t_{0}\cos k_{i}$, the cubic moments above reproduce the weak-field limit of
$t_{1}^{i}$, confirming that the asymmetry is carried by the $k^{4}$ lattice term.
It should be noted that for this nearest-neighbor band $\xi^{\prime\prime}$ only shifts the center of each cosine band. 
Such a common shift is a static Peierls phase, which can be removed by a gauge transformation in a uniform bulk and becomes physical only relative to an undriven momentum reference, such as an interface or the edge of the illuminated region.

We next restore the altermagnetic order
and first consider the $d_{x^{2}-y^{2}}$ case
[Figs.\,\ref{figure2}(c,d)].
Here the kinetic and magnetic terms reside on the same axial bonds and acquire identical bond phases, so that $M^{\prime\prime}$ displaces both spin bands by the same $(\delta_{x},-\delta_{y})$ [Fig.\,\ref{figure2}(c)].
The spin dependence of the Fermi surfaces in Fig.\,\ref{figure2}(d) therefore originates from the even part $M^{\prime}$: the anisotropic dressing $t_{J}^{Cx}\neq t_{J}^{Cy}$ breaks $[C_{2}\|C_{4}^{z}]$ and generates a uniform spin splitting.
For $\beta=0$ the two spin Fermi surfaces are strongly unequal and shifted along $k_{x}$, whereas for $\beta=\pi/4$ the common displacement lies along $k_{x}=-k_{y}$ and the two Fermi surfaces are much closer in size.
To make the hierarchy explicit, we expand
$M_{\mathrm{eff}}=M^{\prime}+M^{\prime\prime}$ around $\Gamma$ up to quadratic order,
\begin{align}
M^{(d)}_{\mathrm{eff}}(\bm{k})
\simeq&
\underbrace{M_{s}^{(0)}+M_{s}^{(2)}k^{2}}_{M_{s}}
+\underbrace{M_{p}^{x}k_{x}+M_{p}^{y}k_{y}}_{M_{p}}\cr
+&\underbrace{M_{d}^{C}(k_{x}^{2}-k_{y}^{2})
+M_{d}^{S}k_{x}k_{y}}_{M_{d}},
\label{Mhierarchy}
\end{align}
with $M_{s}^{(0)}=-2(t_{J}^{Cx}-t_{J}^{Cy})-t_{J}^{S-}$,
$M_{s}^{(2)}=(t_{J}^{Cx}-t_{J}^{Cy}+t_{J}^{S-})/2$,
$M_{p}^{x}=-2t_{J'}^{Cx}-t_{J'}^{S+}$,
$M_{p}^{y}=-2t_{J'}^{Cy}-t_{J'}^{S-}$,
$M_{d}^{C}=(t_{J}^{Cx}+t_{J}^{Cy})/2$, and $M_{d}^{S}=-t_{J}^{S+}$.
Eq.\,(\ref{Mhierarchy}) reveals a complete induced $s$-$p$-$d$ hierarchy, whose components map directly onto the two ingredients. The constant $s$-wave term $M_{s}^{(0)}$ originates from the
anisotropy of the drive and appears already for a single linearly polarized light \cite{Fu2025Light,Fu2025Floquet}, even in the continuum model. 
In contrast, all light-induced terms of order $k^{2}$, i.e., the dressing of the $d$-wave component and the isotropic correction
$M_{s}^{(2)}k^{2}$, are absent in the continuum model, where the $k^{2}$ coefficient keeps its static value; they arise from the higher-gradient lattice terms and require no frequency mixing.
Most importantly, the $p$-wave component requires both ingredients: like the induced $k^{2}$ terms it arises from the lattice, and in addition it requires the odd moment of the even-ratio drive, so that it vanishes for either light alone. The continuum model thus misses both the induced $q=2$ and the $q-1=1$ components, which the lattice supplies.

Let us now consider how this differs for a $d_{xy}$ altermagnet [Figs.\,\ref{figure2}(e,f)].
Here the magnetic term resides on the diagonal bonds, which are not shared by the kinetic term and acquire different bond phases. 
Importantly, this leads to a genuinely spin-dependent odd structure that cannot be removed by a common momentum shift.
This is most transparent when the linearly polarized light is aligned with a nodal direction of the $d_{xy}$ spin splitting, e.g., $\beta=0$, and $\phi=\pi/2$.
In this case $A_{2+}=A_{2-}$, so that
$t_{J}^{S-}=t_{J'}^{S-}=0$ and the $s$-wave component vanishes, while $\cos\phi=0$ removes the odd kinetic term $\xi^{\prime\prime}$.
The only odd term left is $-t_{J'}^{S+}\sin k_{x}\cos k_{y}$, a $p_{x}$-wave spin splitting satisfying $H_{\sigma}^{\mathrm{eff}}(k_{x},k_{y})
=H_{-\sigma}^{\mathrm{eff}}(-k_{x},k_{y})$
[Fig.\,\ref{figure2}(f)]. 
In the weak-field, low-momentum limit and for $\eta=1$ and $k_{y}=0$, we have
\begin{equation}
\frac{M^{(d)}_{\mathrm{eff}}(\bm{k})-M^{(d)}_{\mathrm{eff}}(-\bm{k})}{2}
=
-\frac{t_{J}^{S}}{8}A_{1}^{2}A_{2}\,k_{x}
+\mathcal{O}(A^{5},k^{3}),
\label{dxy-odd-check}
\end{equation}
which vanishes for either light alone and is absent from the leading continuum $d_{xy}$ spin splitting $\propto k_{x}k_{y}$. 
Thus, both ingredients are needed to convert the even-parity $d$-wave into an odd-parity $p$-wave component. 
More generally, varying $\beta$ and $\phi$ activates other symmetry-allowed components, resulting in a hybrid $s$-$p$-$d$ spin splitting.

\textit{$f$-wave magnet: An odd-parity case}---
Let us now consider an odd-parity parent, for which light should
convert the parity in the opposite direction, generating
even-parity components from an odd-parity spin splitting.
We take an $f$-wave magnet on a triangular lattice with the spin
splitting \cite{Ezawa2024b,zt4l-y18j}
\begin{equation}
M_{\mathrm{lat}}^{(f)}(\bm{k})
=
2J_f\sin k_x[\cos k_x-\cos(\sqrt{3}k_y)],
\label{Mf-parent-lattice}
\end{equation}
where $J_f$ sets the strength of the order.
In this dimensionless convention, the three bonds, $\bm{R}=(2,0)$ and $(1,\pm\sqrt{3})$, all with $|\bm{R}|=2$, enter Eq.\,(\ref{Mf-parent-lattice}), which then reduces to $-J_f k_x(k_x^2-3k_y^2)+\mathcal{O}(k^5)$ with a leading continuum form of order $q=3$.
Following the same Floquet procedure as in the $d$-wave case and
expanding the light-dressed spin splitting around $\Gamma$ up to
cubic order, while keeping the full light-amplitude dependence,
we obtain
\begin{equation}
M_{\mathrm{eff}}^{(f)}(\bm{k})
\simeq
M_{s}^{(f)}(\bm{k})
+
M_{p}^{(f)}(\bm{k})
+
M_{d}^{(f)}(\bm{k})
+
M_{f}^{(f)}(\bm{k}),
\label{Mf-hierarchy}
\end{equation}
with the explicit components given in the \textbf{End Matter}.

The driven $f$-wave magnet thus develops a complete induced
$s$-$p$-$d$-$f$ hierarchy [Fig.\,\ref{figure1}(c)], whose
components again map onto the two ingredients:
The $p$-wave component has the parity of the parent and appears
already for a single linearly polarized light, even in the continuum model \cite{Fu2025All,Fu2025Floquet}.
Interestingly, the even-parity $s$-wave component also survives in the continuum model, since the cubic moment of the $\mathcal{N}=2$ drive acting on the $k^{3}$ term yields a constant, although it requires both lights.
In contrast, the $d$-wave component is the missing $q-1$ parity and requires both the even-ratio drive and the higher-gradient lattice terms, while the induced $f$-wave component of order $q$ requires the lattice alone.
Remarkably, the drive therefore converts an odd-parity spin splitting into even-parity $s$- and $d$-wave components, opposite to the $d$-wave case. 
The $g$- and $i$-wave cases follow the same rule and are given in the \textbf{End Matter}.

It is important to note that the labels $s$, $p$, $d$, $f$, $\ldots$ used here classify the angular harmonics of the light-dressed spin splitting around the $\Gamma$ point, i.e., the
components varying as $\cos\ell\theta_{\bm{k}}$ or $\sin\ell\theta_{\bm{k}}$ with leading order $k^{\ell}$, where $\theta_{\bm{k}}$ is the polar angle of $\bm{k}$ and $\ell=0,1,2,3,\ldots$ for $s$, $p$, $d$, $f$, $\ldots$.
They do not refer to irreducible representations of the crystalline point group, which is in any case lowered by the
drive. 
For example, the $f$-wave magnet above is defined on a triangular lattice, whose symmetry differs from that of the square-lattice
$d$-wave case; nevertheless, the quadratic harmonics generated around $\Gamma$ are still $d$-wave components in this sense.
The Floquet hierarchy should therefore be understood as a hierarchy of low-energy momentum harmonics of the light-dressed lattice spin splitting.

\textit{Conclusions.}---
We have shown that a bichromatic drive reorganizes the spin splitting of an unconventional magnet into a hierarchy of lower-order components, and that this hierarchy depends on whether light acts on the lattice or on its continuum limit.
In the leading continuum model, light induces components only up to order $q-2$, while the higher-gradient terms of the lattice supply the missing component of order $q-1$ together with a light-induced correction of order $q$.
Two conditions must therefore be met simultaneously: the frequency ratio must be even, so that odd powers of the field survive the time average, and the higher-gradient lattice terms must be retained. We have then explicitly shown the revised hierarchy for $d$- and $f$-wave magnets, representing even- and odd-parity examples. 
Recent spectroscopic characterizations of altermagnets \cite{Ding2024,Liu2024c} suggest that spin- and time-resolved photoemission and ultrafast scanning-probe techniques could directly resolve the light-induced components in the spin-split bands \cite{Jacobsen2026Probing}. 
Building on the present Letter, future research may explore how light-induced mixed-parity spin splittings affect transport, topology, and superconducting proximity effects in unconventional magnets.

\textit{Acknowledgments.}---
P.-H. F. acknowledges financial support from the Fondazione Cariplo under the grant 2023-2594.
S.A.A.G. acknowledges the U.S. Department of Energy, Office of Science, Basic Energy Sciences, under Award No. DE-SC0024942 for the formal developments, the analytical work, and the writing of the manuscript. S.A.A.G acknowledges the hospitality of the Aspen Center for Physics, which is supported by National Science Foundation grant PHY-2210452 and Kavli Institute for Theoretical Physics (KITP) supported by grant NSF PHY-2309135.

\onecolumngrid
\newpage
\begin{center}
\textbf{\textsc{END MATTERS}}
\end{center}
\vspace{1em}
\twocolumngrid

\textit{E1. Odd-frequency-ratio driving of the $d$-wave magnet.---}
For the odd frequency ratio $\mathcal{N}=\omega_{2}/\omega_{1}=3$, the effective Hamiltonian contains only the $\bm{k}$-even component, $H_{\sigma}^{\mathrm{eff}}(\bm{k})
=
\xi^{\prime}(\bm{k})
+
\sigma M^{\prime}(\bm{k})$.
Neglecting the higher-order frequency-mixing terms proportional to $J_{3}(A_{1})J_{1}(A_{2})$, the corresponding even component is
\begin{align}
\xi^{\prime}(\bm{k})
={}&
4t_{0}-\mu
-2\left(
t_{0}^{x(3)}\cos k_x+t_{0}^{y(3)}\cos k_y
\right),
\\
M^{\prime}(\bm{k})
={}&
-2\left(
t_{J}^{Cx(3)}\cos k_x-t_{J}^{Cy(3)}\cos k_y
\right)
\nonumber
\\
&
-t_{J}^{S+(3)}\sin k_x\sin k_y
-t_{J}^{S-(3)}\cos k_x\cos k_y,
\end{align}
where
\begin{align}
t_{0}^{i(3)}
&\simeq
t_0J_0(A_1)J_0(A_{2i}),
\\
t_{J}^{Ci(3)}
&\simeq
t_J^CJ_0(A_1)J_0(A_{2i}),
\\
t_{J}^{S\pm(3)}
&\simeq
\frac{t_J^S}{4}J_0(\sqrt{2}A_1)
\left[
J_0(A_{2-})\pm J_0(A_{2+})
\right].
\end{align}
Thus the $\mathcal{N}=3$ drive can reshape the $\bm{k}$-even magnetic spin-splitting, including the anisotropic and diagonal components, but it cannot generate the $\bm{k}$-odd component present for $\mathcal{N}=2$.
This $\bm{k}$-even character can be understood from the symmetry of the driving trajectory: For $\mathcal{N}=3$, the vector potential
satisfies $\bm{A}(t+T/2)=-\bm{A}(t)$, so that the trajectory in the $A_x$-$A_y$ plane is inversion symmetric, equivalently with
opposite extrema related by $\bm{A}_{+}=-\bm{A}_{-}$.
Consequently, the period-averaged Hamiltonian preserves momentum inversion.
More generally, for every odd $\mathcal{N}$ the half-period antisymmetry forces all odd moments of the drive to vanish and therefore forbids an odd-in-momentum component.
For even $\mathcal{N}$ this constraint is absent, so odd-in-momentum terms are symmetry allowed but can still vanish for special phases or polarization directions.

\textit{E2. Light-dressed $f$-wave magnet.---}
For the parent lattice spin-splitting in Eq.\,(\ref{Mf-parent-lattice}),
define the three exact bond-dressing factors
\begin{equation}
\mathcal F_x
=
\left\langle e^{2iA_x(t)}\right\rangle_T,
\qquad
\mathcal F_{\pm}
=
\left\langle
e^{i[A_x(t)\pm\sqrt{3}A_y(t)]}
\right\rangle_T,
\label{f-wave-bond-factors}
\end{equation}
where $\langle\cdots\rangle_T=T^{-1}\int_0^Tdt\,(\cdots)$, and
write $\mathcal F_\nu=C_\nu+iS_\nu$ for $\nu=x,+,-$.
The time-averaged magnetic field is then exactly
\begin{equation}
\begin{aligned}
M_{\mathrm{eff}}^{(f)}(\bm{k})
=
J_f\Big\{&
\operatorname{Im}\!\left[\mathcal F_x e^{2ik_x}\right]
-
\operatorname{Im}\!\left[\mathcal F_+e^{i(k_x+\sqrt{3}k_y)}\right]
\\
&-
\operatorname{Im}\!\left[\mathcal F_-e^{i(k_x-\sqrt{3}k_y)}\right]
\Big\}.
\end{aligned}
\label{Mf-exact-dressed}
\end{equation}
Expanding Eq.\,(\ref{Mf-exact-dressed}) around $\Gamma$ through
cubic order in momentum gives
\begin{subequations}
\label{Mf-components-exact}
\begin{align}
M_s^{(f)}(\bm{k})
&=
s_0^{(f)}(1-k^2),
\\
M_p^{(f)}(\bm{k})
&=
\left[p_x^{(0)}+p_x^{(2)}k^2\right]k_x
+
\left[p_y^{(0)}+p_y^{(2)}k^2\right]k_y,
\\
M_d^{(f)}(\bm{k})
&=
d_c^{(f)}(k_x^2-k_y^2)
+
d_s^{(f)}(2k_xk_y),
\\
M_f^{(f)}(\bm{k})
&=
f_c^{(f)}k_x(k_x^2-3k_y^2),
\end{align}
\end{subequations}
with
\begin{subequations}
\label{Mf-coefficients-corrected}
\begin{align}
s_0^{(f)}
&=
J_f(S_x-S_+-S_-),
\\
p_x^{(0)}
&=
J_f(2C_x-C_+-C_-),
&
p_y^{(0)}
&=
\sqrt{3}J_f(C_--C_+),
\\
p_x^{(2)}
&=
J_f\!\left[-C_x+\frac{C_++C_-}{2}\right],
&
p_y^{(2)}
&=
\frac{\sqrt{3}J_f}{2}(C_+-C_-),
\\
d_c^{(f)}
&=
-J_f\!\left[S_x+\frac{S_++S_-}{2}\right],
&
d_s^{(f)}
&=
\frac{\sqrt{3}J_f}{2}(S_+-S_-),
\\
f_c^{(f)}
&=
-\frac{J_f}{3}(C_x+C_++C_-).
\end{align}
\end{subequations}
These expressions retain the full dependence on the light
amplitudes and phase through the exact factors in
Eq.\,(\ref{f-wave-bond-factors}); no expansion in the driving
amplitudes is required.
They also pass the zero-drive limit identically:
$\mathcal F_x=\mathcal F_+=\mathcal F_-=1$ gives
$s_0^{(f)}=p_{x,y}^{(0)}=p_{x,y}^{(2)}=d_{c,s}^{(f)}=0$ and
$f_c^{(f)}=-J_f$, so that
$M_{\mathrm{eff}}^{(f)}=-J_fk_x(k_x^2-3k_y^2)+\mathcal O(k^5)$,
without spurious $p$-wave admixture.
For a generic $1{:}2$ bichromatic drive, the three bond factors
are inequivalent and generally complex, so the $s$-, $p$-, $d$-,
and $f$-wave components are all symmetry allowed.

\textit{E3. Light-dressed $g$-wave magnet.---}
Following the same procedure, we next consider a $g$-wave altermagnet on a square lattice \cite{Ezawa2024b,zt4l-y18j}
\begin{equation}
M_{\mathrm{lat}}^{(g)}(\bm{k})
=
J_g\sin k_x\sin k_y\left(\cos k_y-\cos k_x\right),
\label{Mg-parent-lattice}
\end{equation}
whose leading expansion is
$M_{\mathrm{lat}}^{(g)}=(J_g/2)k_xk_y(k_x^2-k_y^2)+\mathcal O(k^6)$.
Applying the same lattice Floquet construction and expanding the
resulting magnetic spin-splitting around the $\Gamma$ point, we obtain
\begin{equation}
M_{\mathrm{eff}}^{(g)}(\bm{k})
\simeq
M_{s}^{(g)}(\bm{k})
+
M_{p}^{(g)}(\bm{k})
+
M_{d}^{(g)}(\bm{k})
+
M_{f}^{(g)}(\bm{k})
+
M_{g}^{(g)}(\bm{k}),
\label{Mg-hierarchy}
\end{equation}
where
\begin{widetext}

\begin{subequations}
\begin{align}
M_{s}^{(g)}(\bm{k})
={}&
J_{0}\!\left(\sqrt{5}A_{1}\right)
\left(
\Sigma_{0x}^{-}
+
\Sigma_{0y}^{-}
\right)
\left(
\frac{1}{4}
-
\frac{5}{16}k^{2}
+
\frac{25}{256}k^{4}
\right).
\label{Mg-s}
\\
M_{p}^{(g)}(\bm{k})
={}&
J_{2}\!\left(\sqrt{5}A_{1}\right)
\left(
-\frac{1}{2}
+
\frac{5}{16}k^{2}
\right)
\left(
\Pi_{x}^{-}
-
2\Pi_{y}^{-}
\right)k_{x}
+
J_{2}\!\left(\sqrt{5}A_{1}\right)
\left(
\frac{1}{2}
-
\frac{5}{16}k^{2}
\right)
\left(
2\Pi_{x}^{+}
+
\Pi_{y}^{+}
\right)k_{y}.
\label{Mg-p}
\\
M_{d}^{(g)}(\bm{k})
={}&
J_{0}\!\left(\sqrt{5}A_{1}\right)
\left(
\frac{3}{16}
-
\frac{5}{64}k^{2}
\right)
\left(
\Sigma_{0x}^{-}
-
\Sigma_{0y}^{-}
\right)
\left(k_{x}^{2}-k_{y}^{2}\right)
+
J_{0}\!\left(\sqrt{5}A_{1}\right)
\left(
\frac{1}{4}
-
\frac{5}{48}k^{2}
\right)
\left(
\Sigma_{0x}^{+}
-
\Sigma_{0y}^{+}
\right)
\left(2k_{x}k_{y}\right).
\label{Mg-d}
\\
M_{f}^{(g)}(\bm{k})
={}&
\frac{J_{2}\!\left(\sqrt{5}A_{1}\right)}{48}
\left(
-11\Pi_{x}^{-}
-
2\Pi_{y}^{-}
\right)
k_{x}\left(k_{x}^{2}-3k_{y}^{2}\right)
+
\frac{J_{2}\!\left(\sqrt{5}A_{1}\right)}{48}
\left(
2\Pi_{x}^{+}
-
11\Pi_{y}^{+}
\right)
k_{y}\left(3k_{x}^{2}-k_{y}^{2}\right).
\label{Mg-f}
\\
M_{g}^{(g)}(\bm{k})
={}&
-\frac{7J_{0}\!\left(\sqrt{5}A_{1}\right)}{768}
\left(
\Sigma_{0x}^{-}
+
\Sigma_{0y}^{-}
\right)
\left(
k_{x}^{4}
-
6k_{x}^{2}k_{y}^{2}
+
k_{y}^{4}
\right)
+
\frac{J_{0}\!\left(\sqrt{5}A_{1}\right)}{32}
\left(
\Sigma_{0x}^{+}
+
\Sigma_{0y}^{+}
\right)
\left[
4k_{x}k_{y}
\left(k_{x}^{2}-k_{y}^{2}\right)
\right].
\label{Mg-g}
\end{align}
\end{subequations}
where we define
\begin{subequations}
\label{g-wave-Sigma-definitions}
\begin{align}
\Sigma_{0x}^{\pm}
={}&
J_{0}\!\left(A_{2x}-2A_{2y}\right)
\pm
J_{0}\!\left(A_{2x}+2A_{2y}\right),
\quad
\Sigma_{0y}^{\pm}
=
J_{0}\!\left(2A_{2x}+A_{2y}\right)
\pm
J_{0}\!\left(2A_{2x}-A_{2y}\right).
\\
\Pi_{x}^{\pm}
={}&
J_{1}\!\left(A_{2x}-2A_{2y}\right)
\cos\!\left(2\eta\theta_{A}+\phi\right)
\pm
J_{1}\!\left(A_{2x}+2A_{2y}\right)
\cos\!\left(2\eta\theta_{A}-\phi\right),
\\
\Pi_{y}^{\pm}
={}&
J_{1}\!\left(2A_{2x}+A_{2y}\right)
\cos\!\left(2\eta\theta_{A}+\phi\right)
\pm
J_{1}\!\left(2A_{2x}-A_{2y}\right)
\cos\!\left(2\eta\theta_{A}-\phi\right).
\end{align}
\end{subequations}
Here $\theta_A=\arctan(1/2)$ is the angle of the $(2,1)$ bond
with the $k_x$ axis, while $\eta=\pm1$ is the helicity defined in
Eq.\,(\ref{Afields}).
\end{widetext}

The two-frequency field therefore generates the complete $s$-$p$-$d$-$f$-$g$ hierarchy from the parent $g$-wave magnetic spin-splitting.
Particularly notable are the emergent $p$- and $f$-wave components, which carry odd momentum parity despite the even-parity parent $g$-wave order.
Taken together with the driven $d$-wave case, the $f$- and $g$-wave results demonstrate a systematic Floquet hierarchy: a parent unconventional magnetic spin-splitting can generate all lower-order magnetic components in the low-energy expansion around the $\Gamma$ point.
The simultaneous appearance of opposite-parity components further
shows that the two-frequency drive enables magnetic-parity
conversion in both directions, providing a general route toward
Floquet-engineered mixed-parity unconventional magnetism.

The same order-counting argument extends to an $i$-wave parent.
We do not evaluate a specific $i$-wave lattice model here;
instead, we state the hierarchy expected for a generic lattice
realization.
Accordingly, around the $\Gamma$ point, one
expects
\begin{equation}
M_{\mathrm{eff}}^{(i)}(\bm{k})
\simeq
\sum_{\ell=s,p,d,f,g,h,i}
M_{\ell}^{(i)}(\bm{k}).
\label{Mi-expected}
\end{equation}
In particular, a $h$-wave component is allowed in principle by
the light-dressed lattice spin-splitting, although we do not explicitly
evaluate its coefficient here.
Eq.\,(\ref{Mi-expected}) should therefore be regarded as the
expected extension of the hierarchy established explicitly for
the lower-order parent magnetic spin-splittings.

\end{document}